# Anomalous Magnetic and Thermal Behavior in Some $RMn_2O_5$ Oxides

C. L. Huang[1], C. C. Chou[1], C. J. Ho[1], C. P. Sun[1], B. K. Chaudhuri[2], and H. D. Yang[1*]

[1] Department of physics, National Sun Yat-Sen University, Kaohsiung 804, Taiwan
[2] Department of Solid State Physics, Indian Association for the Cultivation of Science, Kolkata 700032, India

**Abstract.** The $RMn_2O_5$ (R=Pr, Nd, Sm, and Eu) oxides showing magnetoelectric (ME) behavior have been prepared in polycrystalline form by a standard citrate route. The lattice parameters, obtained from the powder XRD analysis, follow the rare-earth contraction indicating the trivalent character of the R ions. Cusp-like anomalies in the magnetic susceptibility curve and sharp peaks in the specific heat were reported at the corresponding temperatures in $RMn_2O_5$ (R=Pr, Nd, Sm, and Eu) indicating the magnetic or electric ordering transitions.

## INTRODUCTION

Magnetoelectric (ME) effects were observed recently in $TbMnO_3$ [1], $RMnO_3$ (R=Eu, Gd, Tb, and Dy) [2], and $TbMn_2O_5$ [3]. The important feature in these samples is that the (anti) ferromagnetic and ferroelectric orderings coexist at low temperature ($T$<100K), which denote as multiferroics. The magnetization can be controlled by electric field and the spontaneous electric polarization can be controlled by external magnetic field showing the potential in technological applications.

From the analysis of the structural data from earlier neutron diffraction for $TbMn_2O_5$ [4], it is quite clear that the interaction between Mn spins and the lattice is very strong. In the present study, we investigate the magnetic and the thermal properties of the $RMn_2O_5$ (R=Pr, Nd, Sm, and Eu) system to explore the relationship between multiferroic transition and lattice effects.

## RESULTS AND DISCUSSION

$RMn_2O_5$ (R=Pr, Nd, Sm, and Eu) were prepared in polycrystalline form by a standard citrate route. The single phase for all samples was confirmed by XRD with $CuK\alpha$ radiation. Lattice parameters basically follow the rare-earth contraction indicating trivalent character of all rare earth ions which is consistent with previous work [5], and are listed in Table 1.

The magnetic susceptibility ($\chi$) of $RMn_2O_5$ (R=Pr, Nd, Sm, and Eu) below $T$=150K is shown in Fig. 1. More than one cusp-like anomaly is observed below $T$=100K in Fig. 1 (a) suggesting multiple magnetic transitions. In addition, a Curie-Weiss behavior is found for all samples at around $T$>100K in Fig. 1 (b), from which values of $\mu_{eff}$ are obtained.

Relative specific heat data of $RMn_2O_5$ (R= Pr, Nd, Sm, and Eu) were taken by ac calorimeter [6]. It is clear that multiple phase transitions occur in these samples. A typical specific heat result for $PrMn_2O_5$ is shown in Fig. 2 (a). Three possible transitions occur at 50.4K, 40.2K, and 27.1K which are also observed at corresponding temperatures in magnetic susceptibility data shown in Fig. 2 (b).

To further verify the nature of these phase transitions, more experimental works, such as dielectric constant, doping, pressure and magnetic field effects are in progress and will be published elsewhere [7].

TABLE 1. Lattice parameters, effective moments, and transition temperatures from magnetic susceptibility ($T_M$) and from specific heat ($T_{SH}$) for RMn$_2$O$_5$ (R=Pr, Nd, Sm, and Eu)

| Sample | a(Å) | b(Å) | c(Å) | $\mu_{eff}(\mu_B)$ | $T_M$(K) | $T_{SH}$(K) |
|---|---|---|---|---|---|---|
| Pr | 7.550 | 8.640 | 5.704 | 6.68 | 45.5, 36.8, 26.0 | 50.4, 40.2, 27.1 |
| Nd | 7.500 | 8.616 | 5.699 | 6.88 | 25.9 | 74.3, 58.3, 30.2 |
| Sm | 7.421 | 8.566 | 5.683 | 6.85 | 47.9, 35.7, 21.1 | 49.7, 37.1, 29.9 |
| Eu | 7.382 | 8.548 | 5.680 | 8.67 | 33.3 | 53.2, 43.6, 39.3 |

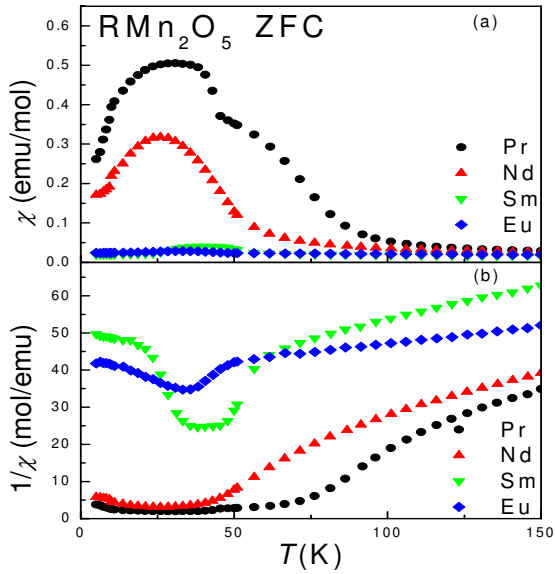

**FIGURE 1.** (a) Magnetic susceptibility ($\chi$) vs. $T$ of RMn$_2$O$_5$ (R=Pr, Nd, Sm, and Eu) measured at applied magnetic field of 1000 Oe. (b) 1/$\chi$ vs. $T$ of RMn$_2$O$_5$ (R=Pr, Nd, Sm, and Eu).

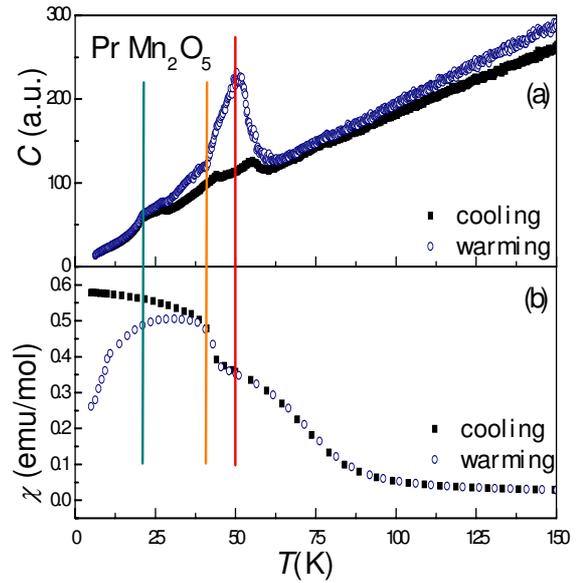

**FIGURE 2.** Temperature dependences of specific heat (a) and magnetic susceptibility (b) for PrMn$_2$O$_5$. Three vertical lines represent the corresponding temperatures where multiple phase transitions may occur.

## CONCLUSION

RMn$_2$O$_5$ (R= Pr, Nd, Sm, and Eu) are prepared and studied with magnetic susceptibility ($\chi$) and specific heat (C) measurements. More than one anomaly is observed in both $\chi$ and C data suggesting that multiple phase transitions, such as ferromagnetic ordering, antiferromagnetic ordering, or ferroelectric ordering occur in this class of materials.

## ACKNOWLEDGMENTS

This research was supported by the National Science Council of the Republic of China under contract Nos. NSC93-2112-M110-001 and NSC93-2112-M009-015.

## REFERENCES

*Corresponding author, email: yang@mail.phys.nsysu.edu.tw

1. T. Kimura *et al.*, *Nature* **426**, 55 (2003).
2. T. Goto *et al.*, *Phys. Rev. Lett.* **92**, 257201 (2004).
3. N. Hur *et al.*, *Nature* **429**, 392 (2004).
4. Buisson G, *Phys. Status Solidi a* **17**, 191 (1973).
5. J. A. Alonso *et al.*, *J. Solid State Chem* **129**, 105 (1997).
6. Y. -K. Kuo *et al.*, *Phys. Rev. B* **64**, 125124 (2001).
7. C. C. Chou *et al.*, unpublished..